\documentclass[aps,pra,twocolumn,amsmath,amssymb,nofootinbib,superscriptaddress]{revtex4}

\newcommand{\perm}[0]{\mathrm{perm}}
\newcommand{\bra}[1]{\langle#1|}
\newcommand{\ket}[1]{|#1\rangle}

\newcommand{\expec}[1]{\langle #1\rangle}

\newcommand{\jonny}[1]{{\color{black}{#1}}}

\DeclareMathOperator{\Var}{Var}
\DeclareMathOperator{\Cov}{Cov}

\usepackage[pdftex]{graphicx}
\usepackage{mathrsfs}
\usepackage[usenames,dvipsnames]{color}
\usepackage{hyperref}
\hypersetup{colorlinks = true, linkcolor = [rgb]{0.19411,0.51882,0.667058}, urlcolor = [rgb]{0.125490,0.29542,0.1647058}, citecolor = [rgb]{0.75882,0.37411,0.14117}}
\usepackage{comment}
\usepackage{amsmath, amssymb}
\usepackage{mathtools}
\usepackage{notes2bib}
\usepackage{float}
\usepackage[normalem]{ulem}
\usepackage{soul}
\setstcolor{red}

\begin{document}

\bibliographystyle{apsrev}

\title{Linear optical quantum metrology with single photons --- \\ Experimental errors, resource counting, and quantum Cram\'er-Rao bounds}

\author{Jonathan P. Olson}
\email[]{olson.jonathanp@gmail.com}
\affiliation{Department of Chemistry and Chemical Biology, Harvard University, Cambridge, Massachusetts 02138, United States}

\author{Keith R. Motes}
\email[]{motesk@gmail.com}
\affiliation{Department of Physics and Astronomy, Macquarie University, Sydney NSW 2113, Australia}

\author{Patrick M. Birchall}
\affiliation{Centre for Quantum Photonics, University of Bristol, Bristol BS8 1UB, UK}

\author{Nick M. Studer}
\affiliation{Hearne Institute for Theoretical Physics and Department of Physics \& Astronomy, Louisiana State University, Baton Rouge, LA 70803, United States}

\author{Margarite LaBorde}
\affiliation{Hearne Institute for Theoretical Physics and Department of Physics \& Astronomy, Louisiana State University, Baton Rouge, LA 70803, United States}

\author{Todd Moulder}
\affiliation{Hearne Institute for Theoretical Physics and Department of Physics \& Astronomy, Louisiana State University, Baton Rouge, LA 70803, United States}

\author{Peter P. Rohde}
\email[]{dr.rohde@gmail.com}
\homepage{http://www.peterrohde.org}
\affiliation{Centre for Quantum Computation and Intelligent Systems, Faculty of Engineering \& Information Technology, University of Technology Sydney, NSW 2007, Australia}
\affiliation{Hearne Institute for Theoretical Physics and Department of Physics \& Astronomy, Louisiana State University, Baton Rouge, LA 70803, United States}

\author{Jonathan P. Dowling}
\affiliation{Hearne Institute for Theoretical Physics and Department of Physics \& Astronomy, Louisiana State University, Baton Rouge, LA 70803, United States}

\date{\today}

\frenchspacing

\begin{abstract}
Quantum number-path entanglement is a resource for super-sensitive quantum metrology and in particular provides for sub-shotnoise or even Heisenberg-limited sensitivity. However, such number-path entanglement has thought to have been resource intensive to create in the first place --- typically requiring either very strong nonlinearities, or nondeterministic preparation schemes with feed-forward, which are difficult to implement. Recently in [Phys. Rev. Lett. \textbf{114}, 170802 (2015)] it was shown that number-path entanglement from a \textsc{BosonSampling} inspired interferometer can be used to beat the shot-noise limit. In this manuscript we compare and contrast different interferometric schemes, discuss resource counting, calculate exact quantum Cram\'er-Rao bounds, and study details of experimental errors. 
\end{abstract}

\maketitle

\section{INTRODUCTION}
A substantial fraction of the efforts of theoretical computer science and physics is now invested in the discovery of post-classical devices to demonstrate quantum supremacy. Much still remains unknown about the fundamental limits and complexity of quantum computing.  One well-known example of a device, which exhibits a quantum advantage over its classical counterpart, came about as a result of the discovery of the Hong-Ou-Mandel effect, enabling interferometers to estimate unknown variables with improved sensitivity \cite{bib:HOM87}.  It was subsequently shown that bosonic NOON states could achieve asymptotically better sensitivities, than a comparable device using only classical techniques, with an increasing number of probe photons $N$ \cite{bib:dowling2008quantum}.  Like universal quantum computers, however, at present the experimental overhead for producing NOON states is prohibitive, making practical implementation infeasible \cite{bib:Kok05, bib:cable2007}.  And although much is now known about metrology using two-mode interferometers, far less is known about larger multi-mode networks \cite{bib:Rafal15}.

Recently, the advent of post-classical devices, like \textsc{BosonSampling} \cite{bib:AA10, bib:Chapter}, has drawn new interest to the capabilities of potentially more practical passive linear-optical networks.  These networks generate complicated number-mode entanglement across an exponentially large state space.  If such an optical network is fed with uncorrelated single photons, the output probabilities in the multimode Fock basis are given by complex matrix permanents, known to be \#P-hard to compute exactly, and strongly believed to be computationally intractable quantities to estimate accurately \cite{bib:Scheel04perm, bib:valiant1979}. Already, \textsc{BosonSampling} has attracted much experimental interest as a simple approach for performing the first truly post-classical computation \cite{bib:Broome20122012, bib:Crespi3, bib:Tillmann4, bib:Spring2, bib:he16}. 

Recently, the sensitivity of a passive linear optical setup, consisting of single photons fed into a specific multi-mode interferometer and photodetection at the output, was investigated \cite{bib:Mordor2015,bib:drummond}. In this manuscript, we  study a much larger class of devices, and further, show how the sensitivity can be maximized in this scenario given realistic constraints on the unknown phase. We show that the device achieving this optimality is not only more sensitive than the one proposed in Ref.~\cite{bib:Mordor2015}, but also far easier to construct. We achieve this by using the Fisher information (FI) formalism, which provides insight into the role of each different component of the interferometer and aids an explicit calculation of the phase sensitivity for the optimal network---a result that was only postulated in Ref.~\cite{bib:Mordor2015}. Additionally, we provide an analytic calculation of the phase sensitivity for the optimal network from matrix permanents, which is an improvement on the conjectured result of Ref.~\cite{bib:Mordor2015}. Our main result shows that for $n<7$ photons we achieve sub-shot-noise limited sensitivity with a passive multi-mode linear optical device with $O(n)$ optical elements. We believe that this work is experimentally achievable with the current temporal infrastructure from Ref.~\cite{bib:he16}.

\jonny{Although previous models for supersensitive quantum devices have shown better theoretical scaling in the limit of larger average photon number, many of these schemes admit a more pessimistic scalability from an engineering perspective.  These devices generally are not robust enough under noise models, require nonlinear components, have a large overhead in state preparation, or employ measurement schemes that are difficult to implement \cite{bib:demkowicz2009,bib:rafalguta}.  For instance, an MZI with a two-mode squeezed vacuum input and parity detection performs extremely well in the noiseless regime, but degrades quickly under dephasing and loss, and requires high efficiency number resolving detectors whose ranges must scale with the average photon number \cite{bib:squeezed}.}

This manuscript is organized as follows: First, in Sec.~\ref{sec:architecture} we describe our generalized architecture, and discuss quantum Fisher information and quantum Cram\'er-Rao bounds. Second, in Sec.~\ref{sec:phasestrat}, we investigate the choice of measurement interferometer in relation to the state preparation interferometer. In this section we also discuss various phase strategies for $\hat{\Phi}$ and find the optimal phase strategy. Third, in Sec.~\ref{sec:OptQuFTI}, we describe how different choices of $\hat{V}_1$ and $\hat{V}_2$ affect device sensitivity and find their optimal structure. Finally, in Sec.~\ref{sec:ExpErr}, we investigate how loss and dephasing errors affect the device performance.

\section{Architecture and Quantum Cram\'er-Rao Bounds} \label{sec:architecture}

\begin{figure}[htb]
\includegraphics[width=0.9\columnwidth]{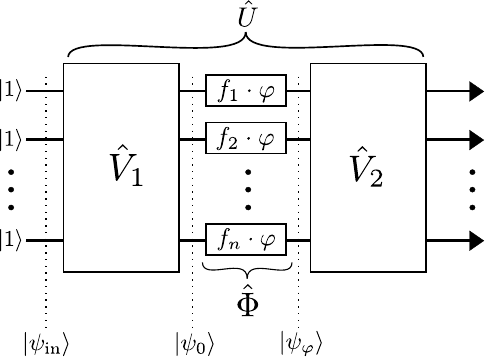}
\caption{A generalized architecture for the quantum Fourier transform interferometer.  We consider optimizations over $V_1,V_2\in$ SU($n$) and phase strategies $\hat{\Phi}$, together with single photon inputs and photodetection in each mode. The MORDOR architecture can be restored when $V_2=V^{\dag}_1$, where $V_1$ is the $n$-mode Quantum Fourier Transform (QFT) and $f_j=(j-1)$ with $j$ being the mode number.}
\label{fig:general}
\end{figure}

To begin, let us consider the Quantum Fourier Transform Interferometer (QuFTI) architecture, as shown in Fig.~\ref{fig:general}, invented by Motes, Olson, Rabeaux, Dowling, Olson and Rohde (MORDOR). A QuFTI consists of three particular components: One, an input state $|\psi_{\text{in}}\rangle=|1\rangle^{\otimes n}$ of $n$ single photons; Two, an $n$-mode interferometer with a transfer matrix $V_1$ which performs the operation $\hat V_1$, followed by a generalized linear phase evolution $\hat \Phi$ which encodes the unknown phase $\varphi$, and a second interference with a transfer matrix $V_2$ (enacting the transformation $\hat V_2$); Three, coincidence photodetection at the output. We will let the phase-evolved state be defined as $\hat \Phi \hat V_1 | \psi_{\text{in}}\rangle \equiv |\psi_{\varphi}\rangle$ where $\hat \Phi = \exp\left(i \sum_j \hat{n}_j f_j \cdot \varphi\right)$, and $\hat n_i$ being the number operator for mode $i$. We define the measurement to be a second evolution with a transfer matrix $V_{2}$ followed by an array of on-off (non-number resolving) photodetectors (see Fig.~\ref{fig:general}) \cite{bib:Mordor2015}. The most compelling aspect of the MORDOR architecture is that it is composed of single-photon inputs, passive linear optics, and on-off detection, all of which can be implemented on an integrated photonic chip  \cite{bib:carolan2015universal}. These simplifications forego many of the technical challenges required to generate NOON states \cite{bib:cable2007}, and is experimentally scalable, \jonny{though only provides sub-shotnoise sensitivity for a small number of modes}. 

The interferometers $\hat{V}_1$, $\hat{V}_2$, and $\hat{\Phi}$ can be varied without jeopardizing the scalability of the device (see Fig.~\ref{fig:general}). Hence, the question arises, what are the optimal choices of $\hat{V}_1$, $\hat{V}_2$ and $\hat{\Phi}$ that yield maximum phase sensitivity? In this work we answer this question. 

To evaluate the phase sensitivities of different architectures, we utilize the quantum Fisher information (QFI) formalism, which we will briefly summarize. Once an unknown parameter $\varphi$ has been encoded onto a quantum state $\rho_{\varphi}$, the QFI $\mathcal{F}(\rho_{\varphi})$ bounds the achievable precision to which $\varphi$ can be estimated, with an unbiased estimator, through the quantum Cram\'er-Rao bound (QCRB)  \cite{bib:helstrom1976},
\begin{align}
 \mathcal{F}(\rho_{\varphi}) \geq 1/\Delta^2\varphi,
\end{align}
where $\Delta^2\varphi$ is the variance in the estimate of $\varphi$.  The QCRB can be saturated when one can measure multiple copies of $\rho_\varphi$ in an optimal basis \cite{bib:braunstein1994}. As such, the QFI is one way to quantify the information content of a quantum state related to an unknown parameter. If $\varphi$ has been encoded into a pure probe state $|\psi_0\rangle$ by a unitary operator, $\ket{\psi_\varphi} = e^{i \hat H \varphi}\ket{\psi_0}$, then the QFI is given by $\mathcal{F}(\ket{\psi_{\varphi}}) = 4(\langle\hat H^2\rangle_0 - \langle \hat H\rangle_0^2)$, where \mbox{$\langle \bullet \rangle_0 = \bra{\psi_0}\bullet\ket{\psi_0}$ \cite{bib:braunstein1994}}.

\section{Phase Strategies} \label{sec:phasestrat}

We begin evaluating the MORDOR framework by investigating the role of $\hat{V}_2$. We note that we measure $\ket{\psi_\varphi}=\hat{\Phi}\hat{V}_1|\psi_{\text{in}}\rangle$ by sending it through the inverse optical network $\hat{V}_2=\hat{V}_1^{\dagger}$ and detecting it using single-photon detectors is an optimal measurement strategy around $\varphi = 0$. This measurement strategy projects $\ket{\psi_\varphi}$ onto a basis containing $\ket{\psi_{\varphi=0}}$. Projecting onto this basis was shown to be an optimal measurement in Ref.~\cite{bib:braunstein1994}. To see this more directly, we can compute the probability $P$ for the observable event $\hat O = (|1\rangle\langle 1|)^{\otimes n}$ of detecting a photon at each output of the full interferometer $\hat{V}_2 \hat{\Phi} \hat{V}_1$ to be $P = 1- \varphi^2 \mathcal{F}(\ket{\psi_{\varphi}})/4 + \mathcal{O}(\varphi^4)$ when $\hat{V}_2=\hat{V}_1^{\dagger}$, as shown in Appendix \ref{sec:qcrb}. This allows us to directly compute the phase sensitivity $\Delta\varphi$ obtained from the error propagation formula in terms of QFI,
\begin{align}
\begin{split}
\Delta\varphi=&\frac{\sqrt{\expec{\hat{O}^2}-\expec{\hat{O}}^2}}{|\frac{d\expec{\hat{O}}}{d\varphi}|} 
\\
=&1/\sqrt{\mathcal{F}(|\psi_{\varphi}\rangle)} + \mathcal{O}(\varphi^2).
\end{split}
 \label{eq:errorprop}
\end{align}
Thus, for $\varphi \simeq 0$, the error propagation formula saturates the QCRB, and hence, the measurement choice of $\hat{V}_2 = \hat{V}_1^\dagger$ is optimal; this is one of the primary results of this manuscript. Note that when $\hat{V}_2=\hat{V}_1^\dagger$, as the precision calculated by the error propagation formula can be obtained by computing the QFI, the problem of maximizing the phase sensitivity of the device is equivalent to maximizing $\mathcal{F}(\ket{\psi_\varphi})$.

We turn now to the choice of $f_i$'s, which is a particular choice of phase strategy as shown in Fig.~\ref{tab:strategies}. As has been noted before, by passing a probe state through an unknown phase shift $k$ times, the effect of the phase shift is magnified \cite{higgins2007entanglement}. If the total phase shift $k\theta$ applied to a mode can be measured to a given precision $\Delta^2(k\theta)=\alpha$, then the precision to which $\theta$ is measured is increased by a factor of $k^2$, so that $\Delta^2 \theta = \alpha/k^2$. This effect can be used to increase the precision of any procedure, including those comprised entirely of classical techniques. Therefore, in order to compare different quantum strategies fairly, we impose the normalization condition,
\begin{align} \label{eq:normalization}
\sum_i f_i = 1,
\end{align}
i.e., we assume they all utilize the same total accumulated phase.  Now we show that, given this normalization condition, the highest QFI can be achieved using a single phase strategy, i.e. $f_j = \delta_{j,1}$. Analogous to the MORDOR framework, the unknown phase is encoded by the unitary, $\hat \Phi = \exp\left(i \sum_j \hat{n}_j f_j \cdot \varphi\right)$. It follows that,
\begin{align}
\begin{split}
\frac{1}{4}\mathcal{F}(|\psi_{\varphi}\rangle) &= \sum_{j,k} f_j f_k \Cov_0(\hat n_j, \hat n_k) \\
&\leq \sum_{j,k} f_j f_k \Var_0(\hat n_h) \\
&=  \Var_0(\hat n_h),
\end{split}
\end{align}
where the covariance is $\Cov_0(A,B) \equiv  \langle  A B \rangle_0 - \langle  A \rangle_0  \langle B \rangle_0$, for commuting operators $A,B$ and mode $h$ is taken to be the one with the largest photon number variance. If the phase shift was only put in mode $h$, then the QFI would simply be $4\Var(\hat n_h)$. We conclude that distributing the phase shift between two or more modes leads to a QFI, which is less than or equal to the QFI obtained if the phase shift is only in mode $h$. If we assume that the variance is largest in mode 1, which can be ensured by the choice of $\hat{V}_1$, the optimal phase distribution is $f_j=\delta_{j,1}$.  
\begin{table}[h]
\begin{tabular}{l l}
Sub-linear & $f^{\text{sub}}_{j}=\sqrt{j} $ \\
Linear & $f^{\text{lin}}_{j}=j-1$ \\
Quadratic & $f^{\text{quad}}_{j}=j^2$ \\
Exponential & $f^{\text{exp}}_{j}=2^j$ \\
Delta  & $f^{\delta}_{j}=\delta_{j,1}$
\end{tabular}
\caption{Functions representing trial phase strategies.  Note that many of the strategies are not normalized to satisfy Eq.~\eqref{eq:normalization}, but can easily be made so by dividing each by $\sum_{j=1}^n f_j$.}
\label{tab:strategies}
\end{table}

\begin{figure}[htb]
\includegraphics[width=1\columnwidth]{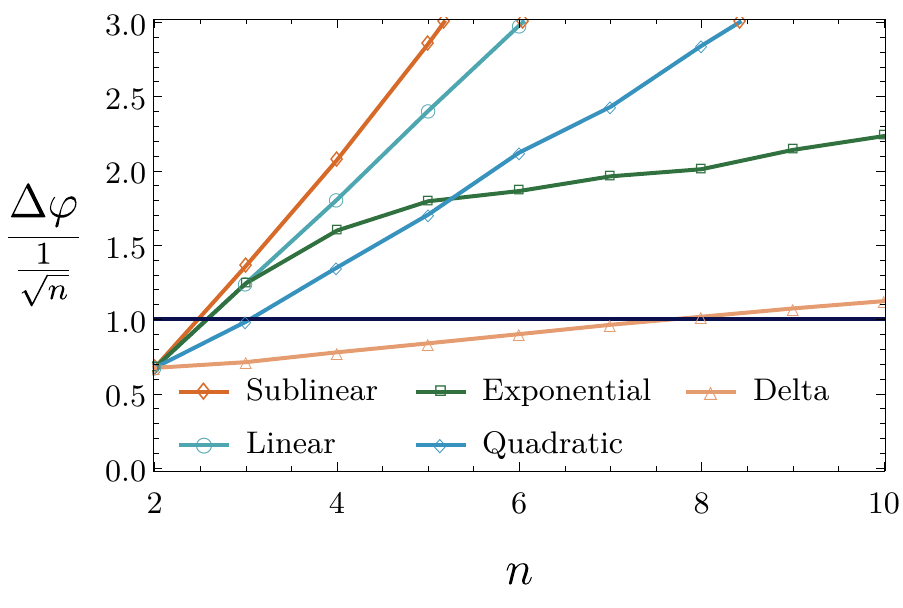}
\caption{The phase sensitivity scaling of different phase strategies for the QuFTI suggests that widening the ``phase gap'' between modes improves the phase sensitivity.  The shot-noise limit used for comparison here is defined to be $1/\sqrt{n}$, which is the best possible classical scheme for $n$ photons and any number of modes greater than two. The region where the sensitivity falls below one indicates super-sensitivity. It is apparent that the delta phase strategy is optimal.}
\label{fig:stratplot}
\end{figure}

To illustrate the performance of differing phase strategies, we consider a range of functions representing trial strategies (Table \ref{tab:strategies}). For each phase strategy, we can use the result of Ref.~\cite{bib:Scheel04perm} to numerically compute the phase sensitivity $\Delta\varphi$ using matrix permanents of $V_1 \Phi V_1^{\dagger}$.  This technique is summarized in Appendix \ref{sec:permintro}, and the result is plotted in Fig.~\ref{fig:stratplot}.  From this figure, it is apparent that there is no improvement in phase sensitivity by distributing the phase throughout the modes, and restricting $\varphi$ to one mode is most effective. This result is consistent with the conclusions by Berry \emph{et al.} as shown in Ref.~\cite{bib:BerryPRA09}. 

With this in mind, we would like to now compare the architecture described in the original MORDOR work to the new phase strategy described above that optimizes the QuFTI. However, we have already made this comparison, since MORDOR possesses the linear phase strategy $f_j=j-1$, whose normalized strategy is plotted against the optimal strategy in Fig.~\ref{fig:stratplot}.  This is contradictory to the \jonny{preliminary results} in MORDOR, which showed that for all $n$, the phase sensitivity of MORDOR beats the shotnoise limit.  This is because MORDOR used a different resource counting technique called \textit{ordinal resource counting} (ORC). The ORC strategy found in MORDOR did not obey the normalization condition on the phase shifts that we have imposed here in Eq.~\eqref{eq:normalization}. For that reason, the comparison in MORDOR with the classical strategy chosen to represent the shot noise limit was unfair.  \jonny{The subsequent errata of Ref.~\cite{bib:Mordor2015} is indeed consistent with the normalized result here.} Thus, when the normalization condition of Eq.~\eqref{eq:normalization} is imposed, the linear phase strategy used in MORDOR is sub-optimal. 

In Fig.~\ref{fig:optimal} we show the phase sensitivity of the QuFTI with the delta phase strategy and compare it to the shot-noise limit and the Heisenberg limit. We see that we do better than shot-noise for $n\leq6$ photons, which is well above what is experimentally achievable today, suggesting that this type of quantum metrology might be the best route forward in the medium-term.

\begin{figure}[htb]
\includegraphics[width=1\columnwidth]{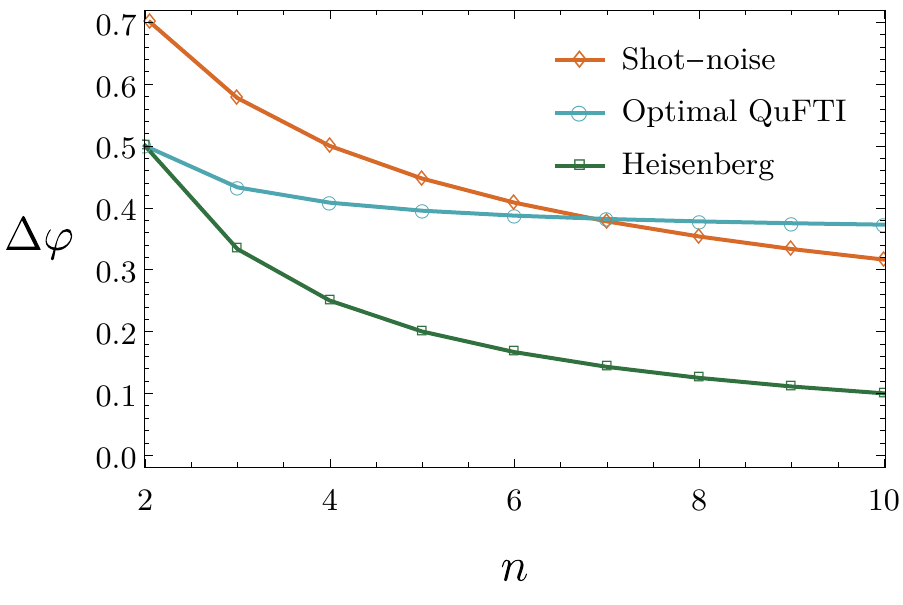}
\caption{Phase sensitivity of the optimal QuFTI, which consists of the delta phase strategy for $\hat{\Phi}$, $\hat{V}_2=\hat{V}^{\dag}_1$, where $\hat{V}_1$ is the $n$-mode Quantum Fourier Transform (QFT), compared to the shot-noise limit and the Heisenberg limit.}
\label{fig:optimal}
\end{figure}

\section{The Optimal Unitary} \label{sec:OptQuFTI} 

In this section, we investigate the effect that $\hat{V}_1$ has on the phase sensitivity, when using the previously found optimal components $\hat{V}_2=\hat{V}_1^\dagger$ and $f_i = \delta_{1,i}$, the best phase strategy. In Sec.~\ref{sec:phasestrat} it was shown that when the phase shift is placed in the first mode i.e. $f_i = \delta_{1,i}$, then $\mathcal{F}(\ket{\psi_{\varphi}})  = 4\cdot\Var_{0}(\hat n_1)$. For an initial state of single photons $|1\rangle^{\otimes n}$ fed into $\hat{V}_1$, we can explicitly compute, as shown in Appendix \ref{sec:holland}, that $\mathcal{F}(\ket{\psi_{\varphi}}) = 8\left(1-\sum_{i=1}^n|V_{1,i}|^4\right)$, where $V_{1,i}$ is the $i^{th}$ element in the top row of $V_1$. Physically, this means the Fisher information is dependent only on the coupling between the input modes and and the first output mode.  This is perhaps to be expected, since only the first mode contains the phase to be interrogated. Additionally, we can compute the QFI when $k$ photons $|k\rangle^{\otimes n}$ are fed into each mode of $\hat{V}_1$ to be,
\begin{align}
\mathcal{F}(\ket{\psi_{\varphi}}) = 4\left(1-\sum_{i=1}^n|V_{1,i}|^4\right)k(k+1),
\end{align}
which is maximized for $V_1$ with $|V_{1,i}| = 1/\sqrt{n}$ for all elements on the top row. Note that while the QuFTI satisfies this constraint, it does not do so uniquely.  To be as general as possible, then, we consider any unitary with this structure to be ``uniform'' and any interferometer with these unitaries to be a Quantum Uniform Multi-mode Interferometer (QUMI). A QuFTI is hence a special case of a QUMI. Remarkably, because the phase sensitivity of the device relies only on the values of the first row of the matrix $V_1$, a network can attain the maximum sensitivity with only $O(n)$  beamsplitters---this is a significant improvement over the MORDOR architecture's QuFTI, which requires $O(n^2)$ beamsplitters to implement.  A simple implementation of the new architecture is shown in Fig.~\ref{fig:optimal}, where the reflectivity amplitude of the beamsplitter acting on modes 1 and $k$ should be $1/\sqrt{k}$. 
Setting $|V_{1,i}| = 1/\sqrt{n}$ gives,
\begin{align}
\mathcal{F}_{\mathrm{max}}(\ket{\psi_{\varphi}}) = 4k(k+1)\left(1-1/n\right).
\end{align}
This reduces to the result of Holland and Burnett when $n=2$ modes \cite{bib:holland1993}. Members of our team have shown that it is possible to make $k$-photon Fock input states from only single photons using more advanced devices, such as reliable quantum memory, which is beyond current experimental techniques \cite{bib:motes2016efficient}.  

For an input state of single photons the optimal precision obtainable is therefore,
\begin{align}
\Delta \varphi = 1/\sqrt{8(1-1/n)}. \label{eq:deltaphifinal}
\end{align}
One can also arrive at the same result by an explicit calculation of matrix permanents when $V=V_1=V_2^\dagger$, which we provide in Appendix \ref{sec:Uentries}. \\

\begin{figure}[htb]
\includegraphics[width=0.9\columnwidth]{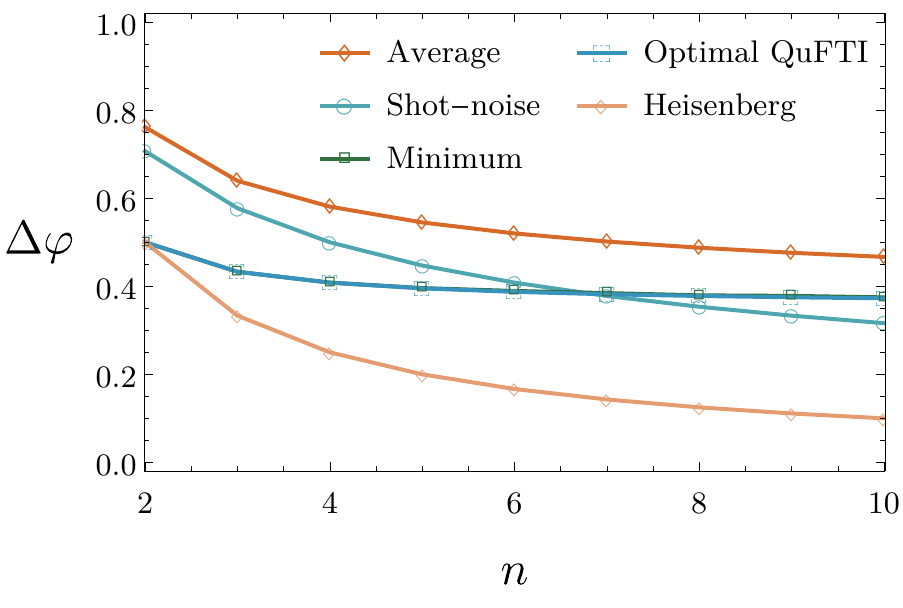}
\caption{The Quantum Fourier Transform (QFT) is optimal for the delta phase strategy.   However, it is not uniquely so -- we know that any uniform unitary is also optimal for this strategy.}
\label{fig:minuplot}
\end{figure}

To investigate the sensitivity of a device in which $\hat{V}_2 = \hat{V}_1^\dagger$ but wherein $\hat{V}_1$ is not optimal, we computed the phase sensitivity of 10,000 random unitaries in SU($n$) (for each $n$), and plotted the best phase sensitivity (i.e. minimum $\Delta\varphi$) and average phase sensitivity of this set against the phase sensitivity of the optimal QUMI (see Fig.~\ref{fig:minuplot}). It is now clear that the best strategy is to use the delta phase function and the QUMI. We call this overall best strategy, which has the optimal delta phase shift, combined with the uniform weighted first row of the unitaries, the optimal QUMI.  Note that, although an experimental implementation of the QuFTI is not optimal for QUMIs, we use the QuFTI for analytic calculations and numerics, since the matrix itself has useful symmetries and still produces the same output statistics as any QUMI.

\begin{figure}[htb]
\includegraphics[width=0.9\columnwidth]{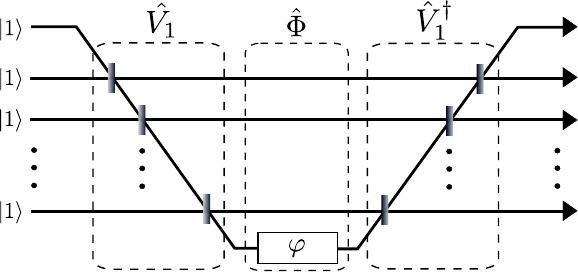}
\caption{A simple architecture which maximizes the phase sensitivity of our scheme.  The beamsplitters (grey boxes) should be adjusted so that $V_1$ is a QUMI, namely, the reflectivity amplitude of the beamsplitter acting on mode 1 and mode $k$ should be $1/\sqrt{k}$.}
\label{fig:optimal}
\end{figure}

\section{Experimental Errors} \label{sec:ExpErr}

Quantum states such as single photons are notoriously difficult to manipulate. It is therefore important to consider how various errors affect the quality of metrology protocols. In particular we look at loss and dephasing.  Particularly, such effects generate mixed states for which the QFI is very difficult to calculate. In addition, the QFI is not practically instructive if you cannot perform an optimal measurement due to technological limitations.

For example, when loss has been applied to a quantum state, the optimal measurement strategies can require exotic techniques such as non-demolition measurement and feed-forward \cite{bib:demkowicz2009}. Therefore calculating the QFI is not necessarily a good indicator of the performance of a practical strategy. Because of these issues, we study the performance of the architecture by directly calculating the sensitivity of specific measurement outcomes from implementable measurement techniques. 

\subsection{Loss Analysis}
\jonny{Loss is a considerable issue to overcome in any experiment utilizing single photons.  
One hurdle when considering photon loss in the scheme presented here is that, if the device relies solely on bucket photon detectors, a loss event is indistinguishable from a photon collision event.  Hence, for small $\varphi$, it may be the case that the loss dominates the number of perceived collision events and one is unable to obtain any useful information about $\varphi$.  Employing limited photo-resolution (i.e. detectors with the capability of distinguishing between one and two or more photons) can partially solve this issue, though photon losses corresponding to one of the photons in a collision event will still continue to degrade the signal.  Of course, the advantage of being able to employ simple photodetectors is lost in the case that you retain the exact architecture previously presented.  However, it is possible to implement pseudo-number-resolving detectors for small numbers of photons by simply coupling the output modes with one or more additional vacuum modes via beamsplitters; the output of these additional modes then can use the simpler photodetector. Members of our team have also done an exhaustive calculation of loss in the context of \textsc{BosonSampling} in \cite{bib:motes2015implementing, bib:fiberloop}, where both spatial and temporal losses are considered.}

\begin{figure}[htb]
\includegraphics[width=.9\columnwidth]{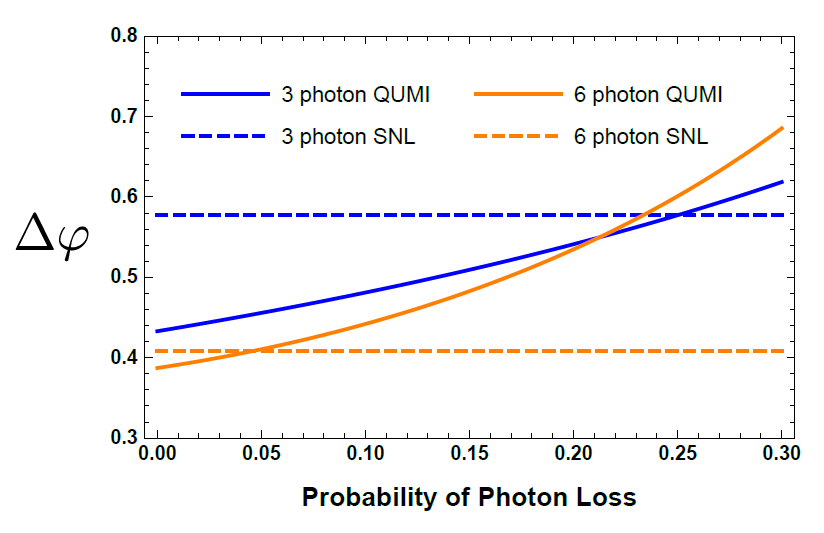}
\caption{Sensitivity of lossy 3- and 6-photon QUMIs (solid lines) compared to their respective shotnoise limits (dashed lines) at $\varphi=0.001$.  The probability on the $x$-axis corresponds to the loss rate for each photon independently in the device.}
\label{fig:loss}
\end{figure}

If enough is known about the error profile beforehand, one can still extrapolate information about $\varphi$ despite the noisy data.  \jonny{For example, if each photon has an independent and equal chance of being lost, then the sensitivity $\Delta\varphi$ degrades continuously according to the photon fidelity $\ell$.  Suppose $P(\varphi)$ corresponds to the probability of measuring the $\ket{1}^{\otimes{n}}$ outcome.  If at most a single photon is lost, the probability of detecting an unambiguous collision event becomes,}
\begin{equation}
\textrm{Pr(collision)}=(1-P)[\ell^n+(n-2)(1-\ell)\ell^{n-1}].
\end{equation}
\jonny{This can then be substituted into the error propagation formula of Eq.~\eqref{eq:errorprop} to give the sensitivity $\Delta\varphi$ (see Fig.~\ref{fig:loss}).}

\jonny{Recent experimental advancements have shown that linear optical networks, particularly for networks of the size considered here, can be constructed with very low loss.  Networks (with even more modes and optical elements than we consider here) have been demonstrated with up to $99\%$ efficiency \cite{effnetwork}; additionally, single photon detectors have achieved up to $93\%$ efficiency  \cite{effdetector}.  Given the rapid advances in engineering optical networks, photon loss may soon be a negligible source of error for the devices considered in this manuscript.}

\subsection{Dephasing Analysis for the Optimal QuFTI} \label{sec:dephasing}

Another type of error often present in optical networks is dephasing. Here we will analyze dephasing in the optimal QuFTI architecture and compare them to NOON states in a standard Mach-Zehnder interferometer.

\begin{figure}[htb]
\includegraphics[width=0.9\columnwidth]{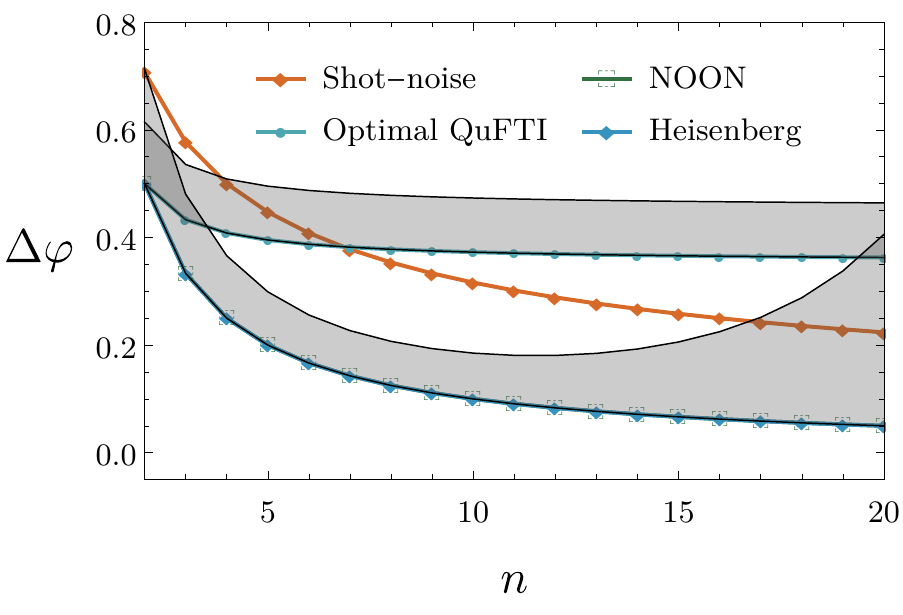}
\caption{Dephasing with the optimal QuFTI for $\varphi=0.1$. This is compared to the shot-noise limit, the Heisenberg limit, and the NOON state with dephasing \cite{bib:Bardhan2013}. The shaded regions represent the dephasing regime of $0\leq\Delta\chi\leq 0.005$ for both the optimal QuFTI and NOON states.} \label{fig:dephasing}
\end{figure}

Dephasing with a single phase shift is modeled with dephasing only occurring on the mode with the unknown phase. In the rest of the interferometer, dephasing can be made very close to zero. We insert a random phase shift $\chi$ to the single mode, which is a Gaussian random variable of zero mean $\expec{\chi}=0$ but nonzero second order moment $\expec{\chi^2}=\Delta\chi$, 
\begin{equation}
e^{\pm i \varphi} \mapsto e^{\pm i (\varphi + \chi)}.
\end{equation}
When both $\varphi,\chi \ll 1$, the approximate form of $P$ derived in Appendix \ref{sec:Uentries} becomes,
\begin{equation}
P= 1-\frac{2n-2}{n}(\varphi+\chi)^2+O(\varphi^4),
\end{equation}
and correspondingly,
\begin{equation}
\expec{P}\approx 1-\frac{(2n-2)(\varphi^2+\Delta\chi)}{n}, \label{eq:expectation}
\end{equation}
which is then substituted into Eq. \eqref{eq:errorprop} to derive the phase sensitivity.  Using this result, we numerically plot the phase sensitivity with dephasing in Fig.~\ref{fig:dephasing} for $0\leq\Delta\chi\leq 0.005$ and $\varphi=0.1$.  We see that, in the sub-shotnoise regime, the optimal QuFTI interferometer is comparable in dephasing with the NOON state in an MZI---another well-known metrological scheme.

It is important to note that when $\Delta\chi$ is close to or larger than the value of $\sqrt{\varphi}$, the estimator $P$ is very poor because it is unable to differentiate between a positive or negative value of $(\varphi+\chi)$.  Indeed, for $\varphi=0, \Delta\chi\neq 0$, the formula in Eq. \eqref{eq:errorprop} does not converge since $\expec{P}_{\varphi=0}\neq 1$.  A straightforward solution is to use a known, controlled phase to shift the average phase far enough away from the peak of $P$ so that, together with the noise, the phase is predominantly positive or negative.

\section{CONCLUSION}

We have considered a variety of different phase and unitary strategies for implementing a passive, single-photon input multi-mode metrological scheme.  We have shown that the optimal architecture for $n$ single photons is a QUMI, which equally couples each mode to a single phase resource in one of the arms of the interferometer, followed by the inverse of the QUMI.  For \mbox{$n<7$}, the sensitivity of the optimal QUMI is sub-shotnoise.  In the asymptotic limit of $n$ photons, the sensitivity approaches a constant.  This limit, however, assumes all $n$ photons are used in a single experiment.  The scheme can always be made to beat the shotnoise limit asymptotically by choosing an architecture with \mbox{$n<7$} modes and repeating the experiment many times.

Perhaps the most attractive feature of our proposed device is that it can be implemented with only passive linear optics, single photon sources, and on-off detectors. In addition, it does not require any of the complicated phase arrangements from the scheme in Motes, Olson, Rabeaux, Dowling, Olson and Rohde. The technology to implement the optimal architecture is essentially identical to that of \textsc{BosonSampling}, and  is achievable with experiments that have already been implemented.

\begin{acknowledgements}
KRM acknowledges the Australian Research Council Centre of Excellence for Engineered Quantum Systems (Project number CE110001013). PPR acknowledges support from Lockheed Martin. JPD acknowledges support from the Air Force Office of Scientific Research, the Army Research Office, The National Science Foundation, and Northrop Grumman Corporation.
\end{acknowledgements}


\appendix

\section{MORDOR Uses Optimal Measurement} \label{sec:qcrb}
Here we link the analysis in MORDOR to the quantum Fisher information formalism and point out that the measurement strategy employed is optimal. To start, we calculate the probability that all photons exit the same mode that they enter,
\begin{align}
\begin{split}
&p(k,\ldots,k) \\
&= \left|\langle k|^{\otimes m}| \hat{V}^{-1}[\mathbb{I}+i\varphi \hat H -\varphi^2\hat H^2/2 -i \mathcal{O}(\varphi^3)]\hat{V}|k\rangle^{\otimes k}\right|^2 \\
&= \left| 1 +i\varphi\langle\hat H\rangle-\varphi^2\langle \hat H^2\rangle/2-i \mathcal{O}(\varphi^3)]\right|^2 \\
&= 1- \varphi^2\big(\langle \hat H \rangle^2-\langle \hat H^2 \rangle\big)+\mathcal{O}(\varphi^4) \\
&= 1- \varphi^2 \mathcal{F}(|\psi\rangle)/4 + \mathcal{O}(\varphi^4),
\end{split}
\end{align}
where we have denoted the unitary performed by the linear optical network as $\hat{V}$. Entering $p(k,\ldots, k)$ into the error propagation formulae,
\begin{align}
\Delta \varphi = \frac{\sqrt{P^2-P}}{\left|\frac{\partial P}{\partial \varphi}\right|},
\end{align}
and evaluating at $\varphi = 0$ gives,
\begin{align}
\Delta \varphi = \frac{1}{\sqrt{\mathcal{F}(|\psi\rangle)}}.
\end{align}
Hence, this measurement basis is optimal as the Quantum Cram\'er-Rao bound is saturated. 

In fact, in a footnote in Ref.~\cite{bib:braunstein1994} it was noted that a measurement strategy, which projected onto the state $\exp(i\varphi \hat H)|\psi\rangle$, would indeed be optimal. This is effectively the measurement, which is being performed in the MORDOR framework making an explicit calculation, as shown here, somewhat redundant. 

\section{Connection Between $\Delta\varphi$ and Matrix Permanents} \label{sec:permintro}
Here, we summarize how matrix permanents can be used to compute the phase sensitivity $\Delta\varphi$ of an interferometer of the form of Fig.~\ref{fig:general}. We wish to numerically compute the probability $P$ of measuring a single photon in each mode, which is the observable $\expec{\hat{O}}$ by which we obtain an estimate of $\varphi$.  Since the input state and measurement is fixed across every strategy, we can use the result of Ref.~\cite{bib:Scheel04perm} to compute $P=|\mathrm{perm}(\hat{U})|^2=|\mathrm{perm}(\hat{V}\hat{\Phi}\hat{V}^{\dagger})|^2$, where perm($\cdot$) refers to the matrix permanent given by the equation,
\begin{align}
\perm(\hat{U})=\sum_{\sigma\in S_{n}} \prod_{i=1}^n u_{i,\sigma(i)}\; ,
\end{align}
$S_n$ being the symmetric group of degree $n$.  The phase sensitivity $\Delta\varphi$ is then found by applying the formula for standard error propagation as in Eq.~\eqref{eq:errorprop}, which can be rewritten in terms of $P$ as,
\begin{align}
\Delta\varphi=\frac{\sqrt{P-P^2}}{|\frac{dP}{d\varphi}|}. \label{eq:peqn}
\end{align}

\section{Optimal State Preparation}\label{sec:holland}
Here we calculate the QFI for the MORDOR setup with $k$ photons entering each mode and the phase shift confined to a single mode. As before, let $\hat{V}=\hat{V}_1=\hat{V}_2^\dagger$. Putting the phase shift in the first mode, the QFI can be calculated using the Heisenberg picture. Labelling the modes in between $\hat{V}$ and $V^{\dagger}$ as $b_i$, where $i=1,2,\ldots m$, then the generator of the phase shift is $\hat H= \hat b_{1}^{\dagger} \hat b_{1}$. By utilising $\hat b_i^{\dagger} = \sum_j V_{i,j} \hat a_{j}^{\dagger} $ \cite{vogel2006quantum}, it is clear that only the top row of $V$ is important. Now we evaluate,
\begin{align}
\begin{split}
\langle\psi| \hat b_{1}^{\dagger} &\hat b_{1} \hat b_{1}^{\dagger} \hat b_{1}|\psi\rangle \\
&=\sum_{i,j,q,l} V_{1,i} \bar V_{1,j}V_{1,q}\bar V_{1,l}  \langle k |^{\otimes m} \hat a_{i}^{\dagger} \hat a_{j}  \hat a_{q}^{\dagger} \hat a_{l} |k\rangle^{\otimes m} \\
& =\sum_{q,l, q\neq l} |V_{1,q}|^2| V_{1,l}|^2  \langle k |^{\otimes m} \hat a_{l}^{\dagger} \hat a_{q}  \hat a_{q}^{\dagger} \hat a_{l} |k\rangle^{\otimes m} \\
& \hspace{20pt} + \sum_{q,l } |V_{1,q}|^2| V_{1,l}|^2  \langle k |^{\otimes m} \hat a_{l}^{\dagger} \hat a_{l}  \hat a_{q}^{\dagger} \hat a_{q} |k\rangle^{\otimes m} \\
& = \sum_{q,l,q \neq l} |V_{1,q}|^2|\bar V_{1,l}|^2 k(k+1) +  k^2 \\
& = \sum_{ q}\big(1- |V_{1,q}|^2 \big)|\bar V_{1,q}|^2 k(k+1) +  k^2 \\
& = \bigg(1- \sum_{ q}|V_{1,q}|^4 \bigg) k(k+1) + k^2, \\
\end{split}
\end{align}
and similarly,
\begin{align}
\langle\psi| \hat b_{1}^{\dagger} \hat b_{1} |\psi\rangle
&=\sum_{q,l} V_{1,q}\bar V_{1,l}  \langle k |^{\otimes m}  \hat a_{q}^{\dagger} \hat a_{l} |k\rangle^{\otimes m}\nonumber \\
&=\sum_{q} | V_{1,q}|^2  \langle k |^{\otimes m}   \hat a_{q}^{\dagger} \hat a_{q} |k\rangle^{\otimes m} \\
&= k. \nonumber
\end{align}
The QFI is,
\begin{align}
4 \bigg(1-\sum_{ q} |V_{1,q}|^4 \bigg) k(k+1).
\end{align}
So to maximise the QFI, $\sum_{ q} |V_{1,q}|^4$ should be minimized, which is achieved for $|V_{1,q}| = 1/\sqrt{m}$ giving a QFI of,
\begin{align}
4 \big(1- 1/m) k(k+1).
\end{align}
When the number of modes equals two this reduces to the case studied by Holland and Burnett \cite{bib:holland1993}. We note that the only part of $V$, which played a role in this calculation was the magnitudes of the elements in the top row. Therefore, instead of a QFT circuit, a series of $m-1$ beamsplitters will also be optimal. 

\section{Derivation of $\Delta\varphi$ from Matrix Permanents} \label{sec:Uentries}
In order to derive the analytic form of $\Delta\varphi$ from Eq.~\eqref{eq:peqn}, we first need the matrix entries of the entire network $\hat{U}=\hat{V}\hat{\Phi} \hat{V}^\dagger$. For a single unknown phase shift $\varphi$ in the first mode,  $\hat{\Phi}$ has the matrix form,
\begin{align}
\Phi_{j,k}=\delta_{j,k}(e^{i\varphi})^{\delta_{j,1}}.
\end{align}

Although any choice of uniform $\hat{V}$, such that $|V_{j,1}|=1/\sqrt{n}$, should be optimal for sensitivity, for this derivation we will choose $\hat{V}$ to be the $n$-multi-mode Quantum Fourier Transform Interferometer (QuFTI), which shares this property. The matrix entries of the entire network become,
\begin{align}
U_{j,k}&= (V\Phi V^{\dag})_{j,k} \nonumber \\ 
&= \sum_{l,m=1}^{n}V_{j,l}\Phi_{l,m}V_{m,k}^{\dag} \nonumber \\
&= \sum_{l,m=1}^{n} \underbrace{\frac{1}{\sqrt{n}}\omega_n^{(j-1)(l-1)}}_{V_{j,l}}\underbrace{\delta_{l,m}e^{i\varphi\delta_{l,1}}}_{\Phi_{l,m}}\underbrace{\frac{1}{\sqrt{n}}\omega_n^{(m-1)(1-k)}}_{V_{m,k}^{\dag}} \nonumber 
\end{align}
\begin{align}
&= \frac{1}{n}\Big[e^{i\varphi}+\sum_{l=2}^{n}\omega_{n}^{(j-1)(l-1)}\omega_n^{(l-1)(1-k)}\Big] \nonumber \\
&= \frac{1}{n}\Big[e^{i\varphi}+\sum_{l=2}^{n}(\omega_{n}^{(j-k)})^{(l-1)}\Big] \nonumber \\
&= \frac{1}{n}\Big[ e^{i\varphi}+\sum_{l=1}^{n-1}(\omega_{n}^{(j-k)})^{l}\Big] \label{eq:sum}
\end{align}
\begin{align}
&= 
\begin{cases}
\displaystyle\frac{1}{n}\Big[ e^{i\varphi}+n-1] & j=k \\ \\
\displaystyle\frac{1}{n}\Big[ e^{i\varphi}-1 \Big] & j\neq k 
\end{cases} \nonumber \\
&= \frac{1}{n} \Big[e^{i\varphi}+(\delta_{j,k})n-1 \Big]. \label{eq:entries} 
\end{align}
For $j=k$, it is easy to see the sum in Eq.~\eqref{eq:sum} should be $n-1$ since each term is simply $1^l=1$.  For $j\neq k$, the result follows from the fact that the sum of all $n^{th}$ roots of unity is zero,
\begin{align}
0=\sum_{l=1}^{n}\omega_n^l.\label{eq:sumroots}
\end{align}
The proof for the above follows directly from the geometric series, and it easy to see that it extends to a sum over $\omega_n^{(j-k)}$ as well.

Now that we have the matrix entries of the network, we can compute the permanent of $\hat{U}=\hat{V}\hat{\Phi} \hat{V}^\dagger$ which is, by definition,
\begin{align}
\textrm{perm}(\hat{U}) &= \sum_{\sigma\in S_n}\prod_{j=1}^n \frac{1}{n} \Big[e^{i\varphi}+(\delta_{j,\sigma(j)})n-1 \Big] \nonumber \\
&=  \frac{1}{n^n}\sum_{\sigma\in S_n}\prod_{j=1}^n \Big[e^{i\varphi}+(\delta_{j,\sigma(j)})n-1 \Big]. \label{eq:permsp}
\end{align}
Suppose $\sigma_k$ is some permutation with $k$ fixed points, recalling that a \textit{fixed point} of a permutation $\sigma$ is a value $j\in\{1,..,n\}$ such that $\sigma(j)=j$ (also referred to as a \textit{partial derangement}), then the product $\prod_{j=1}^n$ in Eq.~\eqref{eq:permsp} corresponding to $\sigma_k$ is,
\begin{align}
\prod_{j=1}^n \Big[e^{i\varphi}+(\delta_{j,\sigma_k(j)})n-1 \Big]=[e^{i\varphi}+n-1]^k[e^{i\varphi}-1]^{n-k}.
\end{align}
The sum in Eq.~\eqref{eq:permsp} can now be rewritten in terms of a sum over the number of fixed points in a permutation, whose coefficient $D_{n,k}$ enumerates all permutations in $S_n$ with $k$ fixed points.  The quantities $D_{n,k}$ are referred to as the \textit{rencontres} numbers,
\begin{align}
D_{n,k}=\frac{n!}{k!}\sum_{j=0}^{n-k}\frac{(-1)^j}{j!}.
\end{align}
The permanent is thus,
\begin{align}
\textrm{perm}(\hat{U}) = \frac{1}{n^n}\sum_{k=0}^n D_{n,k} [e^{i\varphi}+n-1]^k[e^{i\varphi}-1]^{n-k}. \label{eq:permfinal}
\end{align}
We are mostly interested in the behavior of $\textrm{perm}(\hat{U})$ for small $\varphi$, where the phase sensitivity is optimal.  To simplify the remaining calculations, we focus our attention on the Taylor expansion of $F_n[\varphi]=\textrm{perm}(\hat{U})$ around the point $\varphi=0$, up to second order,
\begin{align}
F_n[\varphi]\approx F_n[0]+F_n'[0]\varphi+\frac{1}{2}F_n''[0]\varphi^2.\label{eq:taylor}
\end{align}
We can find $F_n[0]$ easily by noting that, because of the product with $[e^{i\varphi}-1]^{n-k}$ the only non-zero term in Eq.~\eqref{eq:permfinal} corresponds to $k=n$,
\begin{align}
F_n[0]=\frac{1}{n^n} D_{n,n}\cdot[1+n-1]^n=1. 
\end{align}
Similarly, the only non-zero terms in $F_n'[0]$ must be derivatives of either $k=n$ or $k=n-1$.  Since $D_{n,n-1}=0$, we need only concern ourselves with the derivative of the $k=n$ term.  Applying the chain rule gives,
\begin{align}
F_n'[0] &= \Bigg[\frac{1}{n^n}D_{n,n}[e^{i\varphi}+n-1]^n\Bigg]'_{\varphi=0} \nonumber \\
&= \Bigg[\frac{1}{n^n}D_{n,n}n[e^{i\varphi}+n-1]^{n-1}ie^{i\varphi}\Bigg]_{\varphi=0} \label{eq:der} \\
&= \Bigg[\frac{1}{n^n}\cdot1\cdot n[1+n-1]^{n-1}\cdot i\Bigg] \\
&=\frac{n^n}{n^n}\cdot i \nonumber \\
&= i.
\end{align}
Evaluating $F_n''[0]$ is only marginally more difficult.  The $k=n$ term can be evaluated by straightforward application of the product rule to Eq.~\eqref{eq:der}.  Also, although the second derivative of the $k=n-2$ term may be non-zero and contains a product, it is only so for the second derivative of $[e^{i\varphi}-1]^{2}$---the other terms originating from the product rule are zero.  Hence, $F_n''[0]$ has only three non-zero terms,
\begin{align}
F_n''[0]&=\Bigg[\frac{1}{n^n}D_{n,n}n[e^{i\varphi}+n-1]^{n-1}ie^{i\varphi}\Bigg]'_{\varphi=0} \nonumber \\
&\quad+\Bigg[\frac{1}{n^n}D_{n,n-2}[e^{i\varphi}+n-1]^{n-2}[e^{i\varphi}-1]^2\Bigg]''_{\varphi=0} \nonumber \\
&=\Bigg[\frac{1}{n^n}D_{n,n}n(n-1)[e^{i\varphi}+n-1]^{n-2}(ie^{i\varphi})^2\Bigg]_{\varphi=0} \nonumber \\
&\quad+\Bigg[\frac{1}{n^n}D_{n,n}n[e^{i\varphi}+n-1]^{n-1}(ie^{i\varphi})^2\Bigg]_{\varphi=0}\nonumber \\
&\quad+\Bigg[\frac{1}{n^n}D_{n,n-2}2[e^{i\varphi}+n-1]^{n-2}(ie^{i\varphi})^2\Bigg]_{\varphi=0} \nonumber \\
&=-\Bigg[\frac{1}{n^n}(n-1)n^{n-1}\Bigg]-\Bigg[\frac{1}{n^n}n^n\Bigg] \nonumber \\
&\quad-\Bigg[\frac{1}{n^n}2D_{n,n-2}n^{n-2}\Bigg] \nonumber \\
&= -\Big[\frac{n-1}{n}+1+\frac{2D_{n,n-2}}{n^2}\Big] \nonumber \\
&= -\Big[\frac{n-1}{n}+1+\frac{n(n-1)}{n^2}\Big] \nonumber \\
&= -\Big[\frac{2n-2}{n}+1\Big] \nonumber \\
&= -\frac{3n-2}{n}.
\end{align}
Thus, Eq.~\eqref{eq:taylor} becomes the simple expression,
\begin{align}
\textrm{perm}(\hat{U})\approx 1+i\varphi-\left(\frac{3n-2}{2n}\right)\varphi^2.
\end{align}
Recall that the probability of observing $n$ photons, each exiting individual ports, is $P=|\textrm{perm}(\hat{U})|^2$.  For small $\varphi$, then,
\begin{align}
P &= \Big|1+i\varphi-\left(\frac{3n-2}{2n}\right)\varphi^2\Big|^2 \nonumber \\
&= \Big(1+i\varphi-\left(\frac{3n-2}{2n}\right)\varphi^2\Big)\Big(1-i\varphi-\left(\frac{3n-2}{2n}\right)\varphi^2\Big) \nonumber \\
&= 1+i\varphi-i\varphi-2\left(\frac{3n-2}{2n}\right)\varphi^2-i^2\varphi^2+O(\varphi^4) \nonumber \\
&= 1-\frac{2n-2}{n}\varphi^2+O(\varphi^4).
\end{align}
Finally, $\Delta\varphi$ becomes,
\begin{align}
\Delta\varphi &= \frac{\sqrt{P-P^2}}{|\frac{\partial P}{\partial \varphi}|} \nonumber \\
&= \frac{\sqrt{1-\frac{2n-2}{n}\varphi^2-1+\frac{4n-4}{n}\varphi^2}}{\frac{4n-4}{n}\varphi} \nonumber \\
&= \frac{\sqrt{\frac{2n-2}{n}\varphi^2}}{2\cdot\frac{2n-2}{n}\varphi} \nonumber \\
&= \frac{1}{2\sqrt{2}\cdot\sqrt{\frac{n-1}{n}}},
\end{align}
which is in agreement with Eq.~\eqref{eq:deltaphifinal}.  The ratio between $\Delta\varphi$ and the shotnoise-limited phase sensitivity for $n$ photons is then,
\begin{align}
\frac{\Delta\varphi}{1/\sqrt{n}}=\sqrt{\frac{n^2}{8(n-1)}},
\end{align}
which is greater than one (i.e. gives an advantage over shotnoise) for $2\leq n \leq 6$. This is more photons than what is experimentally available today.

\end{document}